\documentclass[fleqn,usenatbib]{mnras}

\usepackage{newtxtext,newtxmath}
\usepackage{threeparttable}

\usepackage[T1]{fontenc}
\usepackage{ae,aecompl}

\usepackage{graphicx}	
\usepackage{amsmath}	
\usepackage{amssymb}	
\usepackage{adjustbox}
\usepackage{rotating}
\usepackage{soul}

\title{Abundant Z-cyanomethanimine in the interstellar medium:\\ paving the way to the synthesis of adenine}

\author[V. M. Rivilla et al.]{
V. M. Rivilla$^{1}$\thanks{E-mail: rivilla@arcetri.astro.it},
J. Mart\'in-Pintado$^{2}$, 
I. Jim\'enez-Serra$^{2}$,
S. Zeng$^{3}$,
S. Mart\'in$^{4,5}$,
\newauthor{
J. Armijos-Abenda\~no$^{6}$,
M. A. Requena-Torres$^{7}$,
R. Aladro$^{8}$,
and D. Riquelme$^{8}$ }
\\
$^{1}$INAF-Osservatorio Astrofisico di Arcetri, Largo Enrico Fermi 5, I-50125, Florence, Italy\\
$^{2}$Centro de Astrobiolog\'ia (CSIC$-$INTA). Ctra de Ajalvir, km. 4, Torrej\'on de Ardoz, 28850 Madrid, Spain \\
$^{3}$School of Physics and Astronomy, Queen Mary University of London, Mile End Road, E1 4NS London, United Kingdom\\
$^{4}$ European Southern Observatory (ESO), Alonso de C\'ordova 3107, Vitacura, Santiago, Chile\\
$^{5}$Joint ALMA Observatory, Alonso de C\'ordova 3107, Vitacura, Santiago, Chile\\
$^{6}$Observatorio Astron\'omico de Quito, Escuela Polit\'ecnica Nacional, Av. Gran Colombia S/N, Interior del Parque La Alameda,\\
       170136, Quito, Ecuador \\
$^{7}$University of Maryland, College Park, ND 20742-2421 \\
$^{8}$Max-Planck-Institut f\"ur Radioastronomie, Auf dem H\"ugel 69, 53121 Bonn, 
}

\date{Accepted XXX. Received YYY; in original form ZZZ}

\pubyear{2018}

\begin{document}
\label{firstpage}
\pagerange{\pageref{firstpage}--\pageref{lastpage}}
\maketitle
\raggedbottom

\begin{abstract}
We report the first detection in the interstellar medium of the Z-isomer of cyanomethanimine (HNCHCN), an HCN dimer proposed as precursor of adenine. We identified six transitions of Z-cyanomethanimine, along with five transitions of E-cyanomethanimine, using IRAM 30m observations towards the Galactic Center quiescent molecular cloud G+0.693. The Z-isomer has a column density of (2.0$\pm$0.6)$\times$10$^{14}$ cm$^{-2}$ and an abundance of 1.5$\times$10$^{-9}$. The relative abundance ratio between the isomers is [Z/E]$\sim$6. This value cannot be explained by the two chemical formation routes previously proposed (gas-phase and grain surface), which predicts abundances ratios between 0.9 and 1.5. The observed [Z/E] ratio is in good agreement with thermodynamic equilibrium at the gas kinetic temperature (130$-$210 K). Since isomerization is not possible in the ISM, the two species may be formed at high temperature. New chemical models, including surface chemistry on dust grains and gas-phase reactions, should be explored to explain our findings. Whatever the formation mechanism, the high abundance of Z-HNCHCN shows that precursors of adenine are efficiently formed in the ISM.
\end{abstract}



\section{Introduction}

Understanding the origin of life on Earth is one of the most challenging problems in astrophysics in the framework of astrobiology. To shed light on this complex topic, it is absolutely needed a comprehensive study of the chemical complexity of the interstellar medium (ISM) that feeds the formation of stars and planets. In this sense, the chemical family of nitriles can give important clues. Nitriles, organic compounds with a $-$C$\equiv$N functional group, play a crucial role in prebiotic chemistry since they are key intermediates in the formation of amino acids, peptides, nucleic acids and nucleobases (e.g. \citealt{balucani2009}). Adenine (H$_5$C$_5$N$_5$), one of the nucleobases of DNA and RNA, is a basic ingredient in the RNA-world scenario for the origin of life on Earth (e.g. \citealt{kaiser&balucani2001}, \citealt{ehrenfreund2001}, \citealt{bernstein2004}). \citet{oro1961} reported the synthesis of adenine from a solution of HCN and NH$_3$ under conditions similar to those thought to have existed on the primitive Earth. 
\citet{chakrabarti2000} proposed that adenine might be formed during the chemical evolution of a star-forming molecular cloud through the oligomerization of HCN in the gas phase in four steps:

\begin{equation*}
\rm 
HCN \rightarrow H_2C_2N_2 \rightarrow NH_2CH(CN)_2 \rightarrow 
\end{equation*}
\begin{equation*}
\rm
\rightarrow NH_2(CN)C = C(CN)NH_2 \rightarrow H_5C_5N_5 
\end{equation*}

However, by performing theoretical calculations, \citet{smith2001} and \citet{yim2012} showed that the first step, i.e. the formation of an HCN dimer from two HCN molecules, does not occur efficiently in the conditions of the ISM. 
Therefore, the question of how HCN dimers can be formed remains open. Since they are precursors of adenine, understanding their formation mechanisms is of crucial importance from an astrobiological point of view.

The most stable dimer of HCN is C-cyanomethanimine (HNCHCN hereafter), which presents two different isomers: the Z-isomer and the E-isomer (\citealt{clemmons1983}). These species are stereoisomers about the double bond N=C  (see e.g. Fig. 1 of \citealt{takano1990}) and the conversion from the Z- to the E-isomer requires an energy of 15.95 kK. 
%
The laboratory experiments and chemical calculations by \citet{takano1990} and \citet{zaleski2013}, respectively, indicate that the Z-isomer is more stable and lower in energy than the E-isomer, with an estimated energy difference in the range 238$-$382 K. Nevertheless, only the high-energy E-isomer has been detected in the ISM so far. Several lines were reported in absorption towards the bright continuum of the cluster of hot cores located in the SgrB2N complex (\citealt{zaleski2013}), while the Z-conformer was elusive. Recent searches of the Z-conformer in a sample of low-mass star-forming regions have also been unsuccessful (\citealt{melosso2018}).

In this work, we report the results of an interstellar search for the Z-conformer of HNCHCN (Z-HNCHCN hereafter). Using an IRAM 30m spectral survey, we searched for this species in the G+0.693-0.027 giant molecular cloud (G+0.693 hereafter) in the Galactic Center. This region exhibits high gas kinetic temperatures ranging from $\sim$50$\,$K to $\sim$150$\,$K (\citealt{Guesten1985,huettemeister_kinetic_1993,Rodriguez-Fernandez2001,ginsburg_dense_2016,Krieger2017,zeng2018}), low dust temperatures of $\leq$30$\,$K (\citealt{Rodriguez-Fernandez2004}), and relatively low H$_2$ gas densities ($\sim$10$^4$ cm$^{-3}$; \citealt{rodriguez-fernandez2000}). 
Due to the low H$_2$ densities, the molecules are sub-thermally excited and hence the excitation temperatures are low ($\sim$5-20 K; \citealt{requena-torres_organic_2006,martin_tracing_2008,rivilla2018,zeng2018}).  
G+0.693 is one of the most chemically rich reservoirs in our Galaxy. Many molecular species have been identified in this cloud, including some of prebiotic relevance such as complex organic molecules (COMs;  \citealt{requena-torres_organic_2006,requena-torres_largest_2008,zeng2018}) and phosphorus-bearing species (\citealt{rivilla2018}). In particular, numerous nitrile molecules have been already detected in the source such as C$_3$N, HC$_3$N, HC$_5$N, HC$_7$N, CH$_2$CN, CH$_3$CN, CH$_3$C$_3$N, NH$_2$CN and HOCN (\citealt{zeng2018}). This rich nitrile chemistry makes this source an excellent target to search for more complex nitriles, and in particular Z-HNCHCN.

\begin{table}
\centering
\tabcolsep 2.0pt
\caption{Transitions of HNCHCN detected towards G+0.693.}
\begin{tabular}{c c c c c c}
\hline
Frequency & Transition  & logA$_{\rm ul}$  & E$_{\rm up}$ & $\int{T_{\rm A}^*d}$v & Detection \\
 (GHz) &   &  (s$^{-1}$) & (K) & (K km s$^{-1}$) & level \\
\hline
\multicolumn{6}{c}{Z-isomer} \\
\hline
85.283 & 9(1,9)$-$8(1,8)  & -5.2107  & 23 & 0.68$\pm$0.25 & 13.8 \\
86.996 & 9(0,9)$-$8(0,8)  & -5.1795  & 21 & 0.91$\pm$0.30 & 15.9  \\
89.247 & 9(1,8)$-$8(1,7)  & -5.1515  & 24 & 0.63$\pm$0.24 & 7.7  \\
94.735 & 10(1,10)$-$9(1,9)  & -5.0705  & 27 & 0.47$\pm$0.21 & 13.7  \\
96.569 &  10(0,10)$-$9(0,9)  &  -5.0413 & 26 & 0.62$\pm$0.25 & 7.1  \\
109.017 & 11(1,10)$-$10(1,9)  & -4.8849  & 34 & 0.27$\pm$0.16 & 4.8  \\
\hline
\multicolumn{6}{c}{E-isomer} \\
\hline
84.425  &  9(1,9)$-$8(1,8)  & -4.4608  & 23  & 0.71$\pm$0.05 & 13.9  \\
93.791  &  10(1,10)$-$9(1,9) & -4.3204  &  28  & 0.51$\pm$0.04  & 11.3 \\
95.422  & 10(0,10)$-$9(0,9)  & -4.2937  & 25  & 0.71$\pm$0.05  & 17.9 \\
97.501   & 10(1,9)$-$9(1,8)  & -4.2698  & 29  & 0.47$\pm$0.04 & 8.4  \\
104.895    & 11(0,11)$-$10(0,10)  &   -4.1684 & 30  & 0.46$\pm$0.03 & 6.9 \\
\hline 
\end{tabular}
\label{tab:transitions}
\end{table}


\section{Observations}

We used in this work a spectral line survey towards the quiescent molecular cloud G+0.693 carried out with the IRAM 30m telescope at Pico Veleta\footnote{This paper makes use of observations obtained with the IRAM- 30m telescope. IRAM is supported by INSU/CNRS (France), MPG (Germany), and IGN (Spain).} (Spain) and the NRAO\footnote{The National Radio Astronomy Observatories is a facility of the National Science Foundation, operated under a cooperative agreement by Associated Universities, Inc.} 100$\,$m Robert C. Byrd Green Bank telescope (GBT) in West Virginia (USA), covering frequencies from 12 to 272 GHz . The observations were centered at the coordinates $\alpha$(J2000.0)= 17$^h$ 47$^m$ 22$^s$ and $\delta$(J2000.0)= -28$^{\circ}$ 21$^{\prime}$ 27$^{\prime\prime}$. 
We refer to \citet{zeng2018} for more detailed information on the observations.

\begin{figure*}
\includegraphics[width=15.5cm]{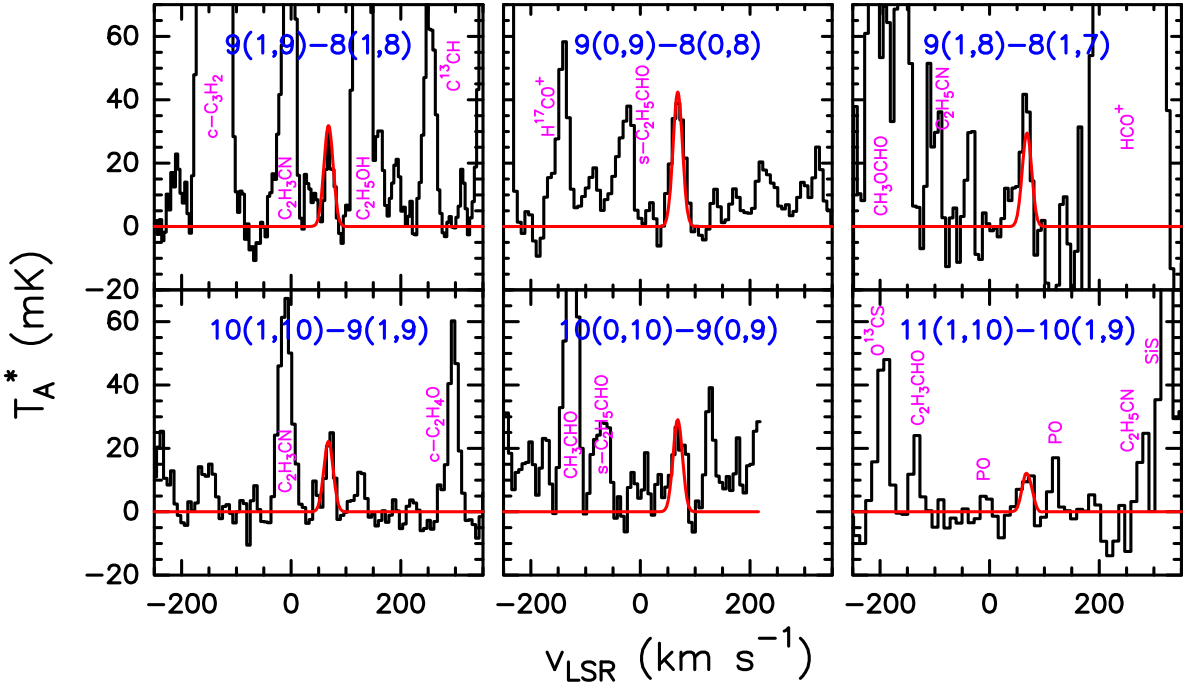}
\caption{IRAM 30m spectra of Z-HNCHCN towards the Galactic Center quiescent giant molecular cloud G+0.693. The red curves correspond to the LTE best fit obtained with MADCUBA$-$AUTOFIT. The quantum numbers of each transition are shown in blue in each panel (see also Table \ref{tab:transitions}). Other molecular species identified in the spectra are indicated with magenta labels.}
 \label{fig-Z}
\end{figure*}

\section{Analysis and Results}
\label{results}

The identification of the lines was performed using the SLIM (Spectral Line Identification and Modeling) tool of the MADCUBA package\footnote{Madrid Data Cube Analysis on ImageJ is a software developed at the Center of Astrobiology (CAB) in Madrid; http://cab.inta-csic.es/madcuba/Portada.html; see Mart\'in et al., in prep; \citet{rivilla_first_2016,rivilla_formation_2017}.}, which contains the spectroscopic information from the Cologne Database for Molecular Spectroscopy (CDMS\footnote{http://www.astro.uni-koeln.de/cdms}) \citep{muller_cologne_2001,muller_cologne_2005,endres_cologne_2016} and the Jet Propulsion Laboratory (JPL) catalogue\footnote{http://spec.jpl.nasa.gov/} \citep{pickett_submillimeter_1998}. 
We have used the CDMS entries for the two isomers of HNCHCN, which contain the spectrocopy from \citet{takano1990},  \citet{zaleski2013} and \citet{melosso2018}. 
Since there are not collisional coefficients available for these species we performed the analysis assuming local thermodynamic equilibrium conditions (LTE). We generated with SLIM-MADCUBA a synthetic spectrum to compare with the observations. 
We confirmed the presence of ten transitions of Z-HNCHCN in the spectra towards G+0.693 with a significant detection level ($>$4$\sigma$; see below). Six of these transitions are unblended, i.e., they are not contaminated by emission from other molecular species, while the other four are blended with other species. We checked that the unblended transitions are not associated with any of the $>$90 species we have identified toward this source (\citealt{requena-torres_largest_2008,rivilla2018,zeng2018}; see Figure \ref{fig-Z}).  
We note that not a single transition of Z-HNCHCN predicted by the LTE spectrum is missing in the data. The spectroscopic information of the six unblended transitions are summarized in Table \ref{tab:transitions}, and the spectra are shown in Figure \ref{fig-Z}. This is the first detection of this species in the ISM.

Then, we used the MADCUBA-AUTOFIT tool that compares the observed spectra with the LTE synthetic spectra, taking into account all the transitions considered, and it provides the best non-linear least-squared fit using the Levenberg-Marquardt algorithm. The free parameters in the fit are: column density ($N$) of the molecule, excitation temperature ($T_{\rm ex}$), velocity ($\varv$), and full width half maximum ($FWHM$). We did not apply a beam dilution factor since it is well known that the molecular emission towards this source is extended over the beam (e.g. \citealt{requena-torres_organic_2006,martin_tracing_2008,rivilla2018}).
For Z-HNCHCN, since the algorithm did not converge leaving all four parameters free, we fixed the linewidth to 20 km s$^{-1}$, which reproduces well the observed spectra and it is consistent with the values inferred for other $-$CN species (\citealt{zeng2018}), and rerun AUTOFIT. The results of the fit are summarized in Table \ref{tab:parameters}. We derived an excitation temperature of 8$\pm$2 K, very similar to that determined for other complex species in this region (\citealt{requena-torres_largest_2008,zeng2018}), and a column density of (2.0$\pm$0.6)$\times$10$^{14}$ cm$^{-2}$. We present in Table \ref{tab:transitions} the velocity integrated intensity ($\int{T_{\rm A}^* d}$v) of the identified transitions resulting from the fit. We calculated the detection level of each transition comparing the velocity integrated intensity with $ rms \times \sqrt{\delta {\rm v}/FWHM}\times FWHM$, where {\it rms} is the noise measured in line-free spectral ranges close to each transition, and $\delta {\rm v}$ is the spectral resolution of the spectra. 
Three of the identified transitions of Z-HNCHCN are above 13$\sigma$, two above 7$\sigma$ and one above 4$\sigma$ (Table \ref{tab:transitions}). 

\begin{figure*}
\includegraphics[width=15.5cm]{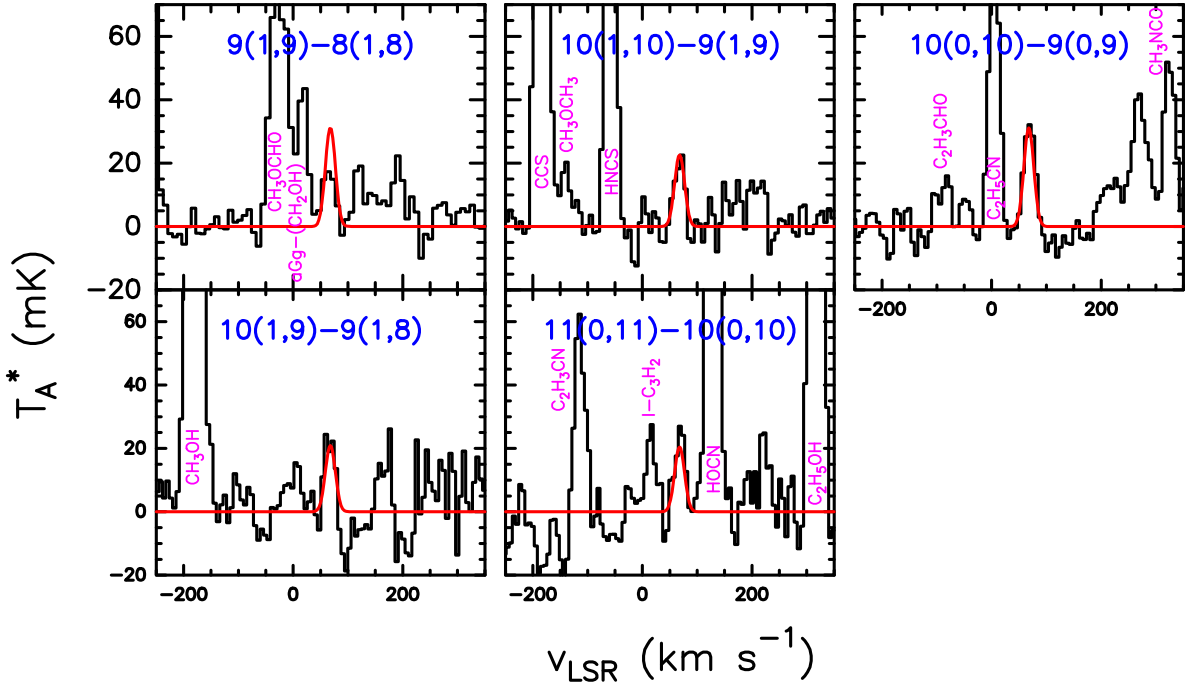}
\caption{IRAM 30m spectra of E-HNCHCN towards the Galactic Center quiescent giant molecular cloud G+0.693. The red curves correspond to the LTE best fit obtained with MADCUBA$-$AUTOFIT. The quantum numbers of each transition are shown in blue in each panel (see also Table \ref{tab:transitions}). Other molecular species identified in the spectra are indicated with magenta labels.}
 \label{fig-E}
\end{figure*}

We repeated the analysis for the E-isomer. We identified eight transitions above $>$4$\sigma$, of which five are unblended  (Table \ref{tab:transitions} and Figure \ref{fig-E}).
Since the AUTOFIT algorithm did not converge leaving T$_{\rm ex}$ as a free parameter, we fixed it to the value found for the Z-isomer, 8 K. We obtained a column density of (0.33$\pm$0.03)$\times$10$^{14}$ cm$^{-2}$. Three transitions are detected above 11$\sigma$, and two above 6$\sigma$ (Table \ref{tab:transitions}).
Both conformers have velocities of around $\sim$68 km s$^{-1}$, consistently with many other molecules observed towards this region (see e.g. \citealt{zeng2018}).
The molecular ratio between the two conformers is [Z/E]=6.1$\pm$2.4.
The total molecular abundance of C-cyanomethanimine, considering both isomers, is 1.74$\times$10$^{-9}$.

We derived the fractional molecular abundances by dividing their column densities by the H$_2$ column density (N$_{\rm H_2}$) measured in G+0.693. We adopted N$_{\rm H_2}$= 1.35 $\times$10$^{23}$ cm$^{-2}$ as inferred by \citet{martin_tracing_2008} from C$^{18}$O observations. In our calculations, we assumed that all molecules show a similar spatial distribution than C$^{18}$O, i.e. all molecules arise from the same volume. The derived abundances are presented in Table \ref{tab:parameters}. The Z-isomer has a relatively high abundance of 1.5$\times$10$^{-9}$, which is comparable to those of other nitrogen-bearing species in this source such as CH$_3$CN or HC$_5$N and higher than those of e.g. CH$_3$NCO and C$_2$H$_5$CN (\citealt{zeng2018}).

We also searched in the spectra of G+0.693 for other molecules that have been proposed as possible precursors of HNCHCN (see further discussion in Section \ref{discussion}): the cyanogen radical (CN), methanimine (CH$_2$NH) and cyanogen (NCCN). Since CN is optically thick towards G+0.693, we have analyzed the optically thin isotopologue $^{13}$CN. The spectra and the spectroscopic information of the studied $^{13}$CN transitions are shown in Appendix \ref{appendix}. The results of AUTOFIT are presented in Table \ref{tab:parameters}. Assuming the isotopic ratio of $^{12}$C/$^{13}$C$\sim$21 derived in this source by \citet{armijos-abendano_3_2014}, we obtained a CN fractional abundance of 1.5$\times$10$^{-8}$. The results of CH$_2$NH were previously presented in \citet{zeng2018} and are also shown in Table \ref{tab:parameters}. 

Since the detection of NCCN is not possible through radio and millimeter observations due to the lack of a permanent electric dipole moment, we searched for its protonated form, NCCNH$^+$. We confirmed the presence of this species through the detection of the J=10-9 and J=11-10 rotational transitions (see Appendix \ref{appendix}). To our knowledge this is the third detection of this species in the ISM after those in the dark cloud TMC-1 and the L483 dense core (\citealt{agundez2015}). Our analysis yielded a fractional molecular abundance of 1.4$\times$10$^{-12}$ for this species (Table \ref{tab:parameters}). If we assume a [NCCNH$^+$]/[NCCN] ratio of $\sim$10$^{-4}$ as inferred from the chemical modelling of \citet{agundez2015}, the abundance of NCCN would be 1.4$\times$10$^{-8}$ cm$^{-2}$.

\section{Discussion}
\label{discussion}

Due to the lack of detections, very little is known about the formation of HNCHCN. There is no chemical formation route of this species included in the astrochemical databases KIDA\footnote{Kinetic Database for Astrochemistry: http://kida.obs.u-bordeaux1.fr} (\citealt{wakelam2012}) and UMIST\footnote{http://udfa.ajmarkwick.net/index.php} (\citealt{mcelroy2013}). 
\citet{zaleski2013} suggested that radical chemistry on the surface of dust grains might form HNCHCN.
More recently, two possible formation routes have been proposed. \citet{vazart2015} studied the neutral-neutral gas-phase reaction between the cyanogen radical and methanimine:

\begin{equation}
\rm
CN + CH_2NH \rightarrow HNCHCN + H
\end{equation}




A different chemical pathway has been proposed by \citet{shivani2017} both on the surface of icy dust grains and in the gas phase:

\begin{equation}
\rm 
NCCN + H +H \rightarrow HNCHCN
\end{equation}

All proposed formation paths seem to be  barrierless, which suggests that both gas-phase and grain surface reactions are able to form efficiently HNCHCN provided that the precursors are sufficiently abundant.
Our data indicate that the reactants of the proposed reactions (CN, CH$_2$NH and NCCN) are relatively abundant in G+0.693, with abundances ranging from 4$\times$10$^{-9}$ to 1.5 $\times$10$^{-8}$ (Table \ref{tab:parameters}), which are higher than the derived abundance of HNCHCN by factor of around 9, 2 and 8, respectively. This suggests that these mechanisms might be able to explain the high abundance of HNCHCN (1.74$\times$10$^{-9}$) inferred in this cloud.

Since we have detected for the first time both isomers, we can use the [Z/E] ratio to constrain the proposed formation scenarios. 
\citet{vazart2015} showed that the gas-phase formation route from CN and CH$_2$NH produces a ratio [Z/E]$\sim$1.5, regardless of the temperature. The calculations by \citet{shivani2017} predict a 
[Z/E] ratio of 0.9 in gas-phase and of 1 on the surface of dust grains. Therefore, both pathways fail to explain the observed ratio of $\sim$6, which might indicate that we are missing key formation routes and/or destruction reactions. A complete study including all the formation and destruction channels of the involved species is needed before drawing firm conclusions. 

Interestingly, the [Z/E] ratio found in G+0.693 seems to indicate that the two isomers are close to thermodynamic equilibrium at the kinetic temperature $T_{\rm k}$ of the cloud. If this is the case, the abundances of the isomers are related through the expression:

\begin{equation}
\label{eq-isomers}
[Z/E]=\frac{N(Z)}{N(E)} = \frac{1}{g} \times exp\left(\frac{\Delta E}{T_{\rm k}}\right) ,
\end{equation}

where $\Delta E$ is the energy difference between the isomers, and $g$ accounts for the statistical weights, which in this case is 1. \citet{takano1990} derived experimentally an energy difference of 237$-$382 K, which is in good agreement with the value of 370 K inferred with the quantum chemical calculations by \citet{zaleski2013}, and with the value of 307 K more recently estimated by \citet{puzzarini2015}.
Then, the observed [Z/E] ratio of 6.1 implies a $T_{\rm k}$ in the range 130$-$210 K, which is in good agreement with the kinetic temperature found by \citet{zeng2018} in G+0.693. This suggests that the two isomers are in thermodynamic equilibrium at the T$_{\rm k}$ of the gas. We note that also the populations of other isomers in the ISM seem to be in thermodynamic equilibrium at T$_{\rm k}$, as e.g. the conformers of ethyl formate (C$_2$H$_5$OCHO) in the hot molecular cores located in the W51 and Orion KL regions (\citealt{rivilla_chemical_2017,tercero_discovery_2013}).
Since the isomerization barrier between the E$-$ and Z$-$isomers of HNCHCN is very high (15.95 kK; \citealt{zaleski2013}) this process cannot occur in the ISM. This means that the $T_{\rm k}$  derived from eq. \ref{eq-isomers} reflects the temperature at which the molecules were formed. Since the dust in G+0.693 is cold ($\leq$30 K; \citealt{Rodriguez-Fernandez2004} ), and the gas temperatures are high ($\sim$50$\,$K to $\sim$150$\,$K;  (e.g. \citealt{Guesten1985,huettemeister_kinetic_1993,Rodriguez-Fernandez2001,ginsburg_dense_2016,Krieger2017,zeng2018}), this opens two possible chemical pathways: 

i) gas-phase reactions occurring at the high kinetic temperatures of the cloud;  and

ii) formation on dust triggered by non-thermal energetic events like cosmic-ray impacts,  and their subsequent release by grain sputtering in moderate-velocity shock waves.
The latter scenario is plausible in the case of G+0.693 since large-scale low-velocity shocks are widespread in the region due to the encounter of two streams of molecular gas (\citealt{hasegawa1994,henshaw2016}). However, the current observations do not allow to discriminate between these two possible chemical routes.

Whatever the formation mechanism, our analysis of the first detection in the ISM of the Z-isomer of HNCHCN reveals that its abundance is higher than that of the E-conformer by a factor of 6. Given the proposed role of HNCHCN as precursor of adenine (\citealt{ESCHENMOSER2007,chakrabarti2000,balucani2012,jung2013}), the relative high abundance of this species, 1.5$\times$10$^{-9}$, argues in favor of an efficient synthesis of key precursors of adenine in space. 
This is a crucial step to understand how the basic ingredients of life could have been assembled in the ISM before their incorporation to the primitive Earth. The role of HNCHCN in the formation of more complex nitrile dimers, and in particular adenine, should be addressed in detail with new detections of HNCHCN in more interstellar sources and with chemical modelling.

\begin{table}
\centering
\tabcolsep 2.5pt
\caption{Derived parameters of the HNCHCN isomers detected towards G+0.693}
\begin{tabular}{l c c c c c}
\hline
Species & N  & T$_{\rm ex}$  & v$_{\rm LSR}$  & FWHM  & Abundance \\
& ($\times$10$^{14}$ cm$^{-2}$) & (K)  &  (km s$^{-1}$) &  (km s$^{-1}$) & ($\times$10$^{-10}$)\\
\hline
Z$-$HNCHCN & 2.0$\pm$0.6  & 8$\pm$2  & 68.3$\pm$0.8 & 20$^{(a)}$ & 15 \\
E$-$HNCHCN & 0.33$\pm$0.03  &  8$^{(a)}$  & 68.0$\pm$0.8 & 21$\pm$2 & 2.4\\
\hline
$^{13}$CN & 0.94$\pm$0.03  & 10$^{(a)}$  & 71.6$\pm$0.4 & 18.8$\pm$0.9 & 7.0 \\
       CN &   &  &  &  & 150$^{(c)}$ \\
CH$_2$NH$^{(b)}$ &  5.4$\pm$0.3 & 9.7$\pm$0.4  & 69$\pm$1  & 25$\pm$1  & 40 \\
NCCNH$^{+}$ & 0.0019$\pm$0.004 & 10$^{(a)}$   & 69$\pm$2 & 22$\pm$5 & 0.014\\
    NCCN &   &  &  &  & 140$^{(d)}$ \\
\hline
\end{tabular}

{(a) Parameter fixed in the MADCUBA$-$AUTOFIT analysis.
(b) From \citet{zeng2018}.
(c) Assuming the isotopic ratio of $^{12}$C/$^{13}$C$\sim$21 derived in G+0.693 by \citet{armijos-abendano_3_2014}.
(d) Assuming a [NCCNH$^+$]/[NCCN] ratio of $\sim$10$^{-4}$, as inferred from chemical modelling by \citet{agundez2015}.
}
\label{tab:parameters}
\end{table}



\section*{Acknowledgements}
We thank the anonymous referee for his/her instructive comments and suggestions. 
V.M.R. has received funding from the European Union's H2020 research and innovation programme under the Marie Sk\l{}odowska-Curie grant agreement No 664931.
J.M.-P. acknowledges partial support by the MINECO and FEDER funding under grants ESP2015-65597-C4-1 and ESP2017-86582-C4-1-R.



\bibliographystyle{mnras}
\bibliography{bibliography} 


\appendix

\newpage

\section{Spectra of $^{13}$CN and NCCNH$^{+}$}
\label{appendix}

\begin{table*}
\centering
\caption{Spectroscopic information from CDMS molecular database of the ransitions of $^{13}$CN detected towards G+0.693.}
\begin{tabular}{c c c c}
\hline
Frequency & Transition  & logA$_{\rm ul}$  & E$_{\rm up}$  \\
 (GHz) &   &  (s$^{-1}$) & (K)   \\
\hline
108.631121 & 1 1 1 0 $-$ 0 1 0 1  &  -5.0188   &  5.2   \\
108.636923 & 1 1 1 1 $-$ 0 1 0 1  &  -5.0172   &  5.2  \\
108.638212 & 1 2 1 1 $-$ 0 1 1 0  &  -5.4444   &  5.2   \\
108.643590 & 1 2 1 2 $-$ 0 1 1 1  &  -5.5926   &  5.2   \\
108.644346 & 1 2 1 0 $-$ 0 1 1 1  &  -5.0188   &  5.2   \\
108.645064 & 1 2 1 1 $-$ 0 1 1 1  &  -5.5620   &  5.2   \\ 
108.651297 & 1 1 1 2 $-$ 0 1 0 1  &  -5.0095   &  5.2   \\
108.657646 & 1 2 1 2 $-$ 0 1 1 2  &  -5.1411   &  5.2   \\
108.658948 & 1 2 1 1 $-$ 0 1 1 2  &  -5.4775   &  5.2   \\  
108.780010 & 1 2 2 3 $-$ 0 1 1 2  &  -4.9788   &  5.2   \\
108.782374 & 1 2 2 2 $-$ 0 1 1 1  &  -5.1107   &  5.2   \\
108.786982 & 1 2 2 1 $-$ 0 1 1 0  &  -5.2430   &  5.2   \\
108.793753 & 1 2 2 1 $-$ 0 1 1 1  &  -5.3502   &  5.2   \\
108.796400 & 1 2 2 2 $-$ 0 1 1 2  &  -5.5604   &  5.2   \\
108.807788 & 1 2 2 1 $-$ 0 1 1 2  &  -6.4897   &  5.2   \\
108.986836 & 1 1 0 1 $-$ 0 1 0 1  &  -7.2578   &  5.2   \\
109.217567 & 1 2 1 2 $-$ 0 1 0 1  &  -6.1603   &  5.2   \\
109.218323 & 1 2 1 0 $-$ 0 1 0 1  &  -6.0475   &  5.2   \\
109.218919 & 1 2 1 1 $-$ 0 1 0 1  &  -6.0932   &  5.2   \\
\hline 
\end{tabular}
\label{tab:13CN}
\end{table*}

\begin{table*}
\centering
\caption{Spectroscopic information from CDMS molecular database of the transitions of NCCNH$^{+}$} detected towards G+0.693.
\begin{tabular}{c c c c}
\hline
Frequency & Transition  & logA$_{\rm ul}$  & E$_{\rm up}$  \\
 (GHz) &   &  (s$^{-1}$) & (K)   \\
\hline
88.758103 &  10$-$9   & -3.7928  &   23  \\
97.633424 &  11$-$10   & -3.6668  &  28   \\
\hline 
\end{tabular}
\label{tab:NCCN+}
\end{table*}

\newpage

\begin{figure*}
\includegraphics[width=14cm]{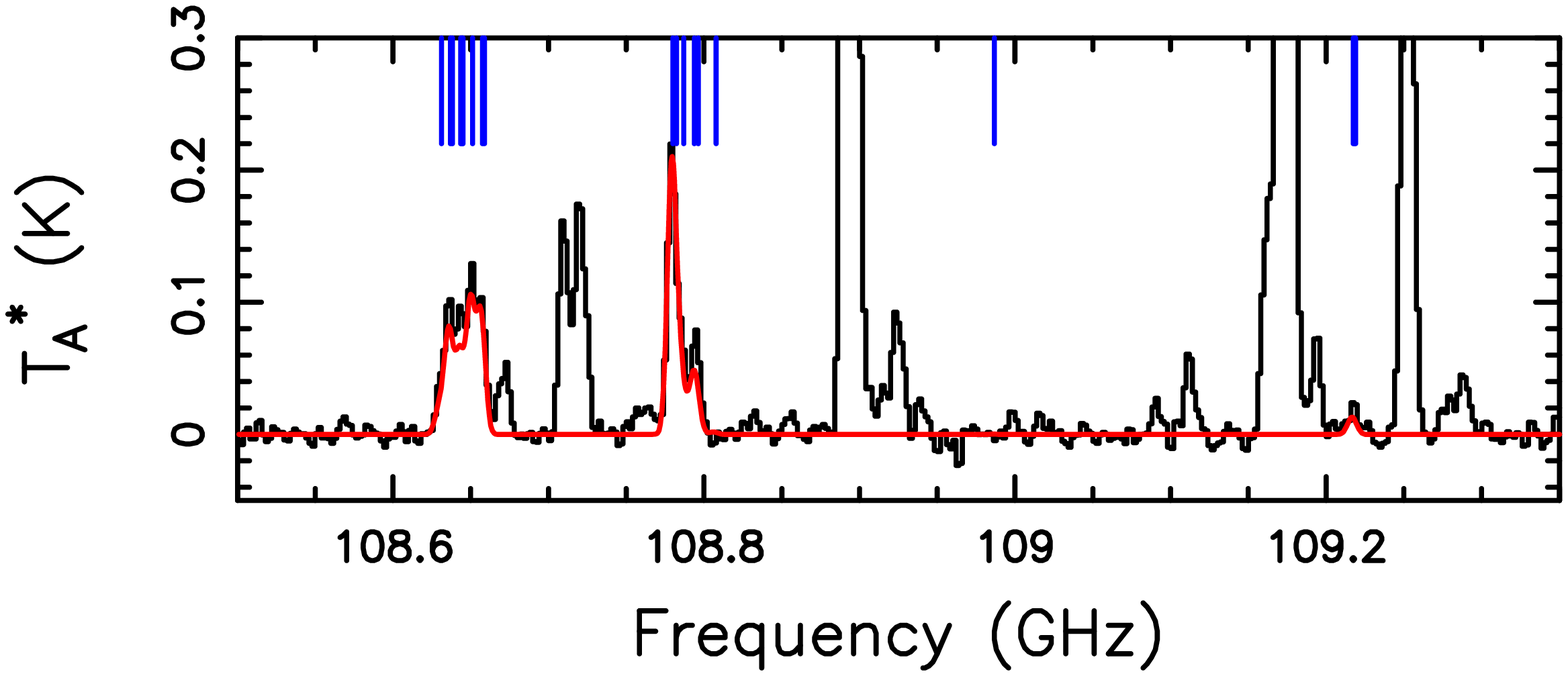}
\caption{IRAM 30m spectra of $^{13}$CN towards the Galactic Center quiescent giant molecular cloud G+0.693. The red curves correspond to the LTE best fit obtained with MADCUBA$-$AUTOFIT. The vertical blue lines indicate the position of the transitions presented in Table \ref{tab:13CN}.}
 \label{fig-other-molecules}
\end{figure*}

\begin{figure*}
\includegraphics[width=14cm]{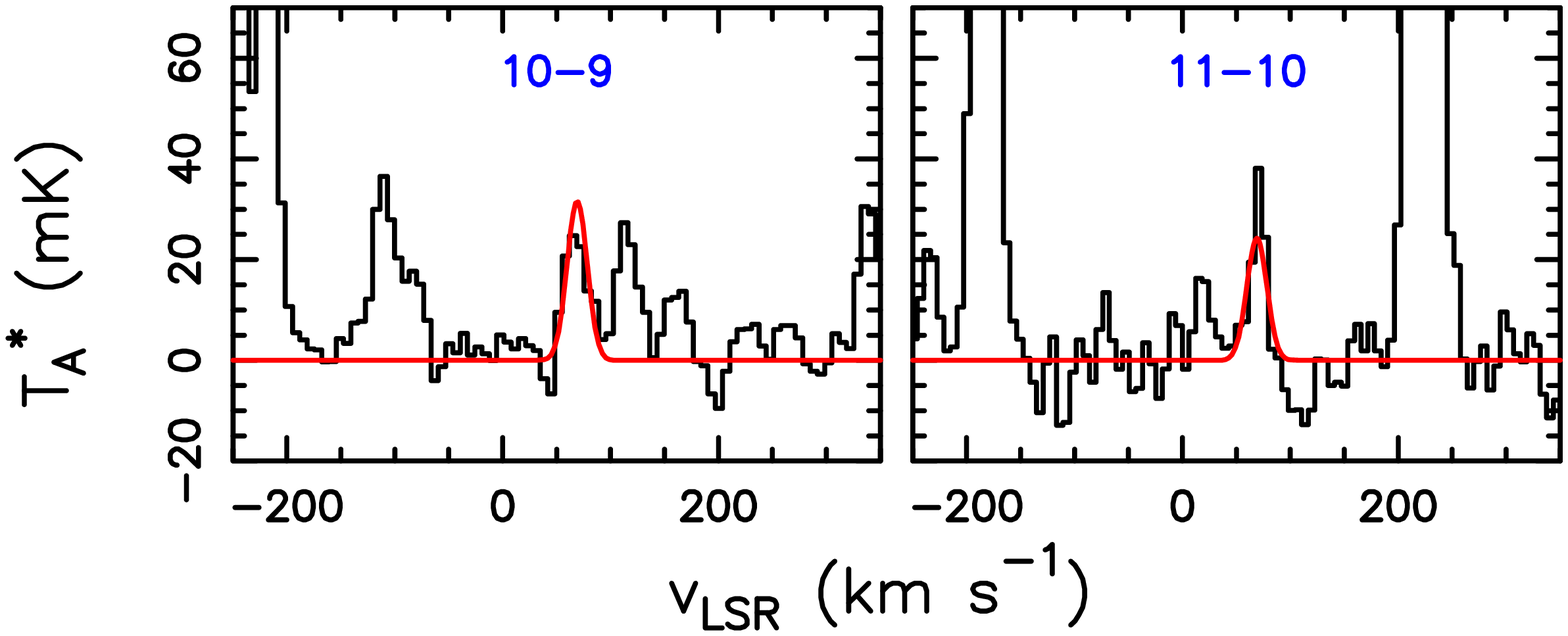}
\caption{IRAM 30m spectra of NCCNH$^{+}$ towards the Galactic Center quiescent giant molecular cloud G+0.693. The red curves correspond to the LTE best fit obtained with MADCUBA$-$AUTOFIT.}
 \label{fig-other-molecules}
\end{figure*}


\bsp	
\label{lastpage}
\end{document}